# Stackelberg game-based optimal scheduling of integrated energy systems considering differences in heat demand across multi-functional areas

Limeng Wang [a], Ranran Yang [b,*], Yang Qu [a], Chengzhe Xu [c]

*a. Key Laboratory of Modern Power System Simulation and Control & Renewable Energy Technology, Ministry of Education （Northeast Electric Power University), Jilin 132012, China；*

*b. State Grid Nanyang Power Supply Company, Nanyang 473000, China*

*c. State Grid Jilin Power Supply Company, Changchun 130000, China*

ABSTRACT

Demand-side management is very critical in China's energy systems because of its high fossil energy consumption and low system energy efficiency. A building shape factor is introduced in describing the architectural characteristics of different functional areas, which are combined with the characteristics of the energy consumed by users to investigate the features of heating load in different functional areas. A Stackelberg game-based optimal scheduling model is proposed for electro-thermal integrated energy systems, which seeks to maximize the revenue of integrated energy operator (IEO) and minimize the cost of users. Here, IEO and users are the Stackelberg game leader and followers, respectively. The leader uses real-time energy prices to guide loads to participate in demand response, while the followers make energy plans based on price feedback. Using the Karush-Kuhn-Tucker (KKT) condition and the big-M method, the model is transformed into a mixed-integer quadratic programming (MIQP) problem, which is solved by using MATLAB and CPLEX software. The results demonstrate that the proposal manages to balance the interests of IEO and users. Furthermore, the heating loads of public and residential areas can be managed separately based on the differences in energy consumption and building shape characteristics, thereby improving the system operational flexibility and promoting renewable energy consumption.

## 1 Introduction

With the increasing energy crisis and environmental pollution, how to build a modern energy system to ensure a safe, clean and sustainable energy supply has become a crucial issue [1]. In this context, vigorously developing and utilizing renewable energies has become an inevitable choice to achieve sustainable energy supply. With the increasing penetration rate of uncertain renewable energy, the secure and stable operation of the system is faced with new challenges [2]. Integrated energy systems may take full advantage of the benefits of multiple complementary energy sources and improve the utilization of renewable energy [3-5]. As a critical component of an integrated energy system (IES) [6,7], electro-thermal integrated energy system can save expenditure on the energy bill via multi-energy complementarity [8]. Deepening the coupling between the power and thermal systems can achieve the effect of energy cascade utilization and improve the comprehensive utilization efficiency, which highlights the importance of collaborative research on the optimal operation of the electric and thermal systems [8]. An electro-thermal integrated energy system may dynamically select resources in two dimensions of time and location to meet the user's power load need and thermal comfort demand [10]. The reason is that the electrical and thermal loads reach their respective peaks at different times [11].

* Corresponding author.
E-mail address: stillwaterrundeep1@163.com.

**Nomenclature**

*Acronyms*

| | |
|---|---|
| IDR | Integrated demand response |
| CHP | Combined heat and power |
| IES | Integrated energy system |
| IEO | Integrated energy operator |
| DHN | District heating network |
| KKT | Karush-Kuhn-Tucker |
| DHS | District heating system |
| PMV | Predicted mean vote |
| PDTP | Dynamic thermal performance |
| BTI | Thermal inertia of buildings |
| DR | Demand response |
| TP | Thermal power |

*Sets and Parameters*

| | |
|---|---|
| $S_{ns}$ | Node set of supply pipes |
| $S_{nr}$ | Node set of return pipes |
| $S_{ps}$ | The pipe sets of supply pipes |
| $S_{pr}$ | The pipe sets of return pipes |
| $I$ | The set of CHP unit |
| $J$ | The set of thermal power unit |
| $U$ | The set of CHP unit, thermal power unit, BESS |
| $Q_{source}$ | The heat energy obtained from heat source (MW) |
| $q_{k,t}$ | The mass flow of pipe k at time $t$ ($kg/s$) |
| $C_{water}$ | Specific heat capacity of hot water medium [$J/(kg \cdot °C)$] |
| $T_{k,t,s}$ | The temperature of water supply pipelines (°C) |
| $T_{k,t,r}$ | The temperature of water return pipelines (°C) |
| $Q_{k,t}$ | The heat energy of the $k$-th pipe at time $t$ (MW) |
| $v_{k,t}$ | The flow velocity of the $k$-th pipe at time $t$ |
| $\tau_{k,t}$ | The delay time of $k$-th pipe (h) |
| $R$ | The average thermal resistance |
| $\Delta Q_k$ | The heat loss of the $k$-th pipe (MW) |
| $T_{soil}^{k,t}$ | The temperature of the soil outside the $k$-th pipe at time $t$ (°C) |
| $H_t$ | The supplied of heat energy at time $t$ (MW) |
| $K_{htc}$ | The heating transfer coefficient of the building |
| $t_m$ | The average temperature of the skin surface (°C) |
| $\pi_l$ | The cost factor for charging and discharging |
| $\kappa_{e/h,\min}$ | Minimum electricity/ thermal price ($/mwh) |
| $\kappa_{e/h,\max}$ | Maximum electricity/ thermal price ($/mwh) |
| $\kappa_{e/h,t,mean}$ | Average electricity and thermal price ($/mwh) |

*Variables*

| | |
|---|---|
| $P_{pv,t}$ | PV power output in period $t$ (MW) |
| $P_{wind,t}$ | WT power output in period $t$ (MW) |
| $P_t^{L,TS}$ | Time-shiftable electric load in period $t$ (MW) |
| $H_{cut,r,t}$ | Cuttable heating load in period $t$ in residential building areas (MW) |
| $H_{cut,p,t}$ | Cuttable heating load in period $t$ in the public building area (MW) |
| $\kappa_{e/h,t}$ | Real-time electricity /thermal price in period $t$ ($/MWh) |
| $P_{con,j,t}$ | Electric power of TP unit $j$ in period $t$ (MW) |
| $P_{chp,i,t}$ | Electric power of CHP unit $i$ in period $t$ (MW) |
| $H_{chp,i,t}$ | Heat power of CHP unit $i$ in period $t$ (MW) |
| $R_{u,t}$ | Reserve capacities of unit $i/j$ or the BESS in period $t$ (MW) |
| $P_{e,cha,t}$ | Charge power of electric storage devices (MW) |
| $P_{e,dis,t}$ | Discharge power of electric storage devices (MW) |
| $P_{h,dis,t}$ | Discharge power of heat storage devices (MW) |
| $P_{h,cha,t}$ | Charge power of heat storage devices (MW) |

In winter, a CHP unit's operation mode is "determining electricity based on heat", which results electricity generated by wind is abandoned [12]. Hence, the researchers introduce electrical and thermal decoupling devices to address the issue. To achieve heat-power decoupling of the CHP unit, electric storage device and heat storage devices are installed next to the CHP unit. Coordinated operation of the electric power system and district heating system, which is an effective way to increase the system operation flexibility, can increase grid-connected electricity from wind power [13-15]. By introducing electro-thermal conversion devices, the fluctuations of renewable energy generations are effectively suppressed, thereby reducing the renewable energy curtailment [16,17]. In reference [18], a model of local scale heat storage is proposed in residential areas. This paper investigates the size of heat storage devices' influence on continuous heating time. Equipment including electric boilers, heat storage devices, heat pumps, etc. is studied. As equipment investment increases, the increment of equipment's heating capacity becomes less ideal.

The traditional dispatching considers the energy balance equation of electrical and thermal loads hourly. Without considering the thermal inertia of thermal networks, the traditional dispatching cannot fully utilize the system's peak shaving ability to promote renewable energy consumption. Making full use of the thermal inertia of buildings and heat network can increase the flexibility of coordinated operation of electric heating systems [19,20]. When the outdoor temperature changes, the heat dissipation process of the building also changes. Because the building has thermal inertia, the building can be regarded as a heat storage device equivalently, which can store and release thermal energy [21]. The demand for heat load is also dynamically changing with the process of heat storage and heat release in the building. To increase the flexibility of the power system, the thermal device responds to the thermal load dispatching within a defined range [22,23].

Several studies have focused on how to take advantage of the thermal characteristics of the district heating system (DHS) to increase system operation flexibility, particularly the dynamic thermal performance (PDTP) of heat network piping and the thermal inertia of buildings (BTI). Considering PDTP means taking into account piping heat loss and thermal time delay, while considering BTI [24] means studying the thermal storage potential of a building as a thermal storage device [25]. In reference [26], a day-ahead optimization scheduling method based on forecasting wind curtailment periods was developed. Reduce the heating load of the building during the wind curtailment period within the adjustable range of the indoor temperature to meet the thermal comfort level. The reduction of heat load can reduce the heat supply of CHP units and promote the absorption of wind power. Because the heating time of all buildings is the same, reference [18] obtains the average heat supply of the building set equivalently from the heat supply of a single building. Then the heat load of the building set is obtained according to the equivalent transformation. Reference [27] establishes a planning model for energy stations in a mixed commercial and residential area, and all nodes are commercial and residential mixed load nodes. In this study, the difference characteristics of electricity and heat loads in mixed commercial and residential areas are relatively lacking. The preceding literature ignores disparities in the thermal storage capacity of different functional types of buildings, as well as differences in the heating demand time of different functional types of buildings.

Demand response is a popular method of transferring energy that helps to maintain the balance between supply and demand [28]. It is regarded as being a crucial part of the development of smart grids in the future. Demand response can reduce the peak-to-valley difference of grid load, which is advantageous to the system's steady operation [29]. Demand response achieves the reasonable distribution of load and energy, which can improve the economy and flexibility of the system [30]. Reference [31] developed a demand-side management technique that adjusts loads in residential areas to reduce energy costs. Reference [32] uses the real-time electricity price to adjust the response quantity of the load, which can reduce the cost and satisfy the user's thermal comfort. A multivariate integrated demand response (IDR) model was built in references [33, 34] guided by electricity price and heat price. A two-stage model of day-ahead and intra-day was established in reference [35]. In the day-ahead stage, the time-shift demand response is used to establish a robust optimization operation model. In the intra-day stage, a replaced DR is used for a stochastic scheduling. This strategy can effectively improve the balance of energy supply and demand by combining the advantages of multiple energy sources. A two-layer optimization model for an integrated energy system was established in reference [36], which took into consideration heating costs and the cost of electricity generation. The nonlinear two-layer model is translated into a mixed-integer linear programming problem using the KKT optimality criterion. The lower-level sub-problem was transformed into the constraints of the upper-level problem. Reference [37] employed Stackelberg game to solve energy management and pricing problems for multi-community integrated energy systems with multi-energy interactions.

Based on existing research, an optimal scheduling approach is presented for integrated energy systems taking into account the differences in heat demand across multiple regions. First, a Stackelberg game-based optimization model is constructed in this paper to condition the interests of integrated energy operator (IEO) and users. To release more thermal flexibility, the temperature threshold of heating buildings with varied functions can be employed to the fullest extent possible. Second, the model is transformed into a mixed-integer quadratic programming (MIQP) using KKT optimum conditions. Last, the simulation results demonstrate that differential heating manages to improve the thermal flexibility of system operation by individually setting the temperature thresholds for buildings in different functional areas.

# 2 Model of electro-thermal integrated energy system

## 2.1 System architecture

An IES is described in this study as a multi-energy system with electro-thermal coupling that includes power producing units, energy conversion devices, and electric storage device. CHP units offer thermal energy, while distributed generator sets and CHP units produce electricity. The electricity grid and heat network are connected by CHP units. Demand response achieves load reduction and shifting. Due to the load peaks of the grid and the heat grid are at different times, the two systems can be connected to improve the energy efficiency of the system. The addition of electric storage devices can shape the electric load curve, cut peaks and fill valleys, and reduce renewable energy curtailment. The introduction of heat storage device facilitates thermal decoupling of CHP units. In Fig. 1, the system architecture is displayed.

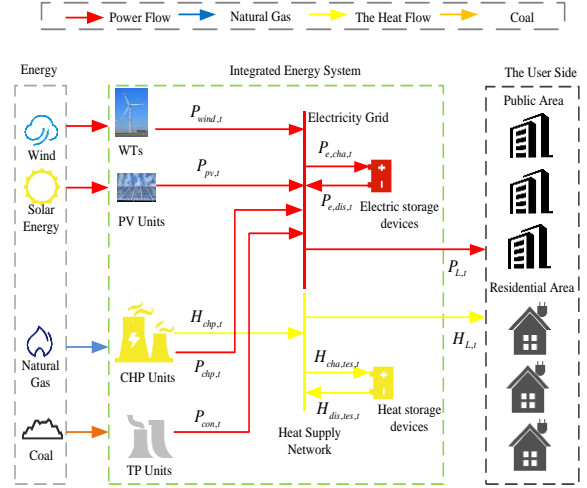

Fig. 1 Architecture of the integrated electric and thermal energy system

As shown in Fig. 1, the IES model used in this study includes CHP units, thermal units, energy storage, wind turbines (WTs), Photovoltaic (PV) units, and loads. Based on the previous work [22], a comparative experiment is set based on different compositions of the auxiliary equipment. In this study, electric and heat storage device is introduced to increase the flexibility of system operation.

## 2.2 Modeling of district heating network
### 2.2.1 Heat source model

Heat source model can be expressed as follows:
$$Q_{source} = q_{k,t} \cdot C_{water} \cdot (T_{k,t,s} - T_{k,t,r}) \quad (1)$$

where $Q_{source}$ is the heat energy obtained from the CHP unit and the heat storage device at the supply network pipe of the head end, $q_{k,t}$ is the mass flow of pipe $k$ at time $t$, $C_{water}$ is specific heat capacity of water, $kJ/(kg \cdot ℃)$, $T_{k,t,s}$ and $T_{k,t,r}$ are the temperatures of water supply or return pipelines at heat source at time $t$, respectively.

### 2.2.2 Mass flow balance of heat-supply network

Similar to Gustav Kirchhoff's law of current, there is a law in heat networks. For each node in the network, the total mass flow in and out of all pipes connected to the node is equal:
$$\sum q_{i',k,t,in} = \sum q_{i',k,t,out} \quad \forall i' \in S_{ns} \cup S_{nr}, \quad \forall k \in S_{ps} \cup S_{pr} \quad (2)$$

where $i'$ is a node of heating networks, $S_{ns}$ and $S_{nr}$ are respectively the node set of supply pipes and return pipes, $k$ is the number of heat-supply pipe, $S_{ps}$ and $S_{pr}$ are the pipe sets of supply/return pipes, respectively. $q_{i',k,t,in}$ is the inlet mass flow of the $k$-th supply/return pipe connecting node $i'$ at time $t$, $q_{i',k,t,out}$ is the outlet mass flow of the $k$-th supply/return pipe connecting node $i'$ at time $t$.

### 2.2.3 Heat power of pipes

The flow of heat medium in a pipe is accompanied by the transfer and loss of heat. According to the basic principle of heat conduction, the heat energy of the heating medium in the pipeline can be calculated according to the following formula:

$$Q_{k,t,in} = q_{k,t} \cdot C_{water} \cdot T_{k,t,in} \quad \forall k \in S_{ps} \cup S_{pr}$$
$$Q_{k,t,out} = q_{k,t} \cdot C_{water} \cdot T_{k,t,out} \quad \forall k \in S_{ps} \cup S_{pr} \quad (3)$$

where $T_{k,t,in}$ and $T_{k,t,out}$ are temperatures of the $k$-th supply/return pipe at time $t$, respectively; $Q_{k,t,in}$ and $Q_{k,t,out}$ are the heat energy of the $k$-th supply/return pipe at time $t$, respectively.

### 2.2.4 The mixed temperature in the node

The temperature of the hot water is fused to the same temperature and delivered to each exit pipe after being gathered from several intake pipes at a single node. The following model may be used to represent the temperature fusion:

$$\sum T_{i',k,t,out} \cdot q_{k,t} = T_{i',t} \sum q_{k,t} \quad \forall i' \in S_{ns} \cup S_{nr}, \\ \forall k \in S_{ps} \cup S_{pr} \quad (4)$$

where $T_{i',k,t,out}$ is the outlet temperature of the $k$-th supply/return pipe connecting node $i'$ at time $t$. $T_{i',t}$ is temperature of supply/return pipe connecting node $i'$ at time $t$.

Assuming the hot water temperature at node $i'$ is fused, the node temperature is equal to the inlet temperature of the pipe starting at node $i'$.

$$T_{i',t} = T_{k,t,in} \quad \forall i' \in S_{ns} \cup S_{nr}, \forall k \in S_{ps} \cup S_{pr} \quad (5)$$

As it travels along the hot water transmission pipeline, the hot water medium absorbs heat from the heat source and transfers it to the load node. Due to the primary heat-supply network's lengthy length and the time delay effect from the supply of heat to the load. The heat-supply network is not a real-time balance system. Therefore, the impact of the hot water flow's time delay on the heat-energy balance of the heat-supply network should be taken into account.

### 2.2.5 Transmission delay of heat-supply network

It is significant to study the transmission delay of heat-supply network in the electric-heating combined dispatching system. The time scale of power system dispatching is very short, while the time scale of heat network transmission delay varies from several minutes to several hours. The heating network satisfies the non-real-time heat balance due to its transmission delay. It increases the scheduling flexibility for the electric heating system and fully utilizes the complementing qualities of different energy sources. This report investigates the major heating network's transmission delay. The principal heating network's hot water transmission rate is as follows:

$$v_{k,t} = \frac{q_{k,t}}{\rho \pi (d_k / 2)^2} \quad \forall k \in S_{ps} \cup S_{pr} \quad (6)$$

where $\rho$ is water density, $d_k$ is diameter of heating network pipe, $v_{k,t}$ is flow velocity of the $k$-th supply/return pipe at time $t$.

The length of the heating pipe and flow velocity are both directly correlated with the time of transmission delay. The time of transmission delay can be expressed as

$$\tau_{k,t} = \sum \frac{L_k}{v_{k,t}} \quad \forall k \in S_{ps} \cup S_{pr} \quad (7)$$

where $\tau_{k,t}$ is actual delay time of $k$-th supply/return pipe.

In order to apply to the optimal scheduling of electro-thermal integrated energy system, the actual delay time is integer by using a rounding function, as shown in the following formula:

$$\tau_{k,t} = round[\tau_{k,t} / \Delta t] \quad \forall k \in S_{ps} \cup S_{pr} \quad (8)$$

where $\tau_{k,t}$ is time delay after rounding.

### 2.2.6 Thermal attenuation

After a certain time delay, the inlet of hot water through the length $L_k$ to the outlet, in this process also accompanied by energy attenuation and time delay. $L_k$ is the length of the heating network pipe. The heat loss of hot medium passing $\Delta Q_k$ through length $L_k$ is shown in the following formula:

$$\Delta Q_k \approx \frac{2\pi(T_{k,t,in} - T_{soil}^{k,t})}{\sum R} L_k \quad (9)$$

where $T_{soil}^{k,t}$ denotes the temperature of the soil outside the $k$-th pipe at time $t$; $T_{k,t,in}$ is the inlet temperature of the $k$-th hot water pipe at time $t$; $R$ stands for average thermal resistance between the heated medium and the surrounding environment.

The heat energy obtained by the load side heat exchanger of the primary heat network is

$$Q_{load,t+\tau} = Q_{source,t} - \sum \Delta Q_K \quad (10)$$

where $Q_{load}$ denotes the heat energy obtained by the load side heat exchanger.

## 2.3 Multi-region thermal demand variability modeling

In winter, around 80% to 90% of the heat energy is supplied to various structures. Due to the structures' ability to store heat, it is widely used to alleviate the problem of wind desertion in the mode of continuous heat and power. Because administrative qualities vary between functional areas, building structure are often diverse. Building structural differences are represented by the body shape coefficient of the structure. The shape coefficients of large-volume buildings are low, small-volume buildings are higher [38], and the public area's shape coefficient is 0.2. The building shape coefficient is 0.4 in residential area.

$$Coefficient = \frac{S}{V} \qquad (11)$$

where *Coefficient* is the building shape factor, $S$ is the external surface area of the building in contact with the outdoor atmosphere, and $V$ is the volume of the building.

Energy supply requirements vary by functional region. A public building's heating load requirement can be altered at different stages. Make sure the interior temperature is higher than the equipment's antifreeze temperature in the night. Allow the indoor temperature to be adjusted according to the needs of the human body during the day. There is no clear time split in the heating demand for residential buildings, necessitating all-weather heating. It is possible to change the heat load within a defined range due to the ambiguity of thermal awareness. The indoor temperature threshold is determined to account for thermal comfort over time.

As a result, the building's temperature and comfort can be maintained while reducing heat supply, making the building's heat load demand more flexible. The thermal transient equilibrium equation for the building can be used to calculate the impact of the amount of heat supplied on the change in indoor temperature [39]:

$$\frac{dT_t^{in}}{dt} = \frac{H_t - (T_t^{in} - T_t^{out}) \cdot K_{htc} \cdot S}{c_{air} \cdot \rho_{air} \cdot V} \qquad (12)$$

where $T_t^{in}$ and $T_t^{out}$ are the indoor and outdoor temperatures at time $t$; $H_t$ is the heat supply, $K_{htc}$, $S$, and $V$ are the heating transfer coefficient of the building, the surface area and volume of the building, respectively. $c_{air}$ and $\rho_{air}$ are the air specific heat capacity and density, respectively.

To determine the link between heating load power and interior temperature, the following equation can be derived:

$$H_t = \frac{(T_t^{in} - T_t^{out}) + \frac{K_{htc} \cdot S}{c_{air} \cdot \rho_{air} \cdot V} \cdot \Delta t \cdot (T_{t-1}^{in} - T_t^{out})}{\frac{1}{K_{htc} \cdot S} + \frac{1}{c_{air} \cdot \rho_{air} \cdot V} \cdot \Delta t} ) \qquad (13)$$

### 2.3.1 Temperature setting for residential heating considering thermal comfort

The heating burden of residents is calculated using the human body's thermal comfort index. The level of occupant comfort with regard to thermal load is assessed using the human thermal comfort index. A person's level of activity changes over time, and so do their temperature requirements. During the night, the body is in a sleeping state and less sensitive to its surroundings. Several thermal comfort and energy demands are sacrificed to adapt the heating strategy to the needs of users at various times. Wind power networks are also provided more space through the reduction of thermal demand.

Human comfort is influenced by physical variables including temperature, wind speed, relative humidity, and others. A straightforward formula is frequently used in engineering practice [40], which leaves out airspeed and air humidity, and is as follows:

$$PMV = 2.43 - \frac{3.76(t_m - t_{in})}{M(I_{cl} + 0.1)} \qquad (14)$$

where $t_m$ is the average temperature of the skin surface, which is 33.5°C, $t_{in}$ is the indoor temperature, $M$ is the metabolic rate of the human body, which is 80 $M/m^2$, and $I_{cl}$ is the thermal resistance of clothing [41].

The human body is not immediately aware of temperature fluctuations when the PMV is in the range of [-0.5, 0.5]. It may also meet the human body's interior temperature requirements in the winter when it changes between [-1, 1]. Set the PMV indication to [-1, 1] for [8:00, 22:00], and [-0.5, 0.5] for the rest of the time ranges.

### 2.3.2 Temperature setting for public area

Different schemes are utilized in public and residential locations. Because public spaces have a more definite use time, they can cut heat supply during non-working hours while maintaining a specified temperature (usually 5°C). As a result, during working hours, the temperature of the public area should be regulated to keep the PMV index within [-0.5, 0.5] and should not be lower than the working temperature during non-working hours.

## 3 Stackelberg game-based electrothermal IES scheduling model

Multilevel programming and game theory are commonly used to coordinate the interests of multiple stakeholders in the optimal scheduling [42, 43]. In this study, a Stackelberg game-optimized scheduling model is developed, where different demand responses for thermal loads are considered. In this model, energy suppliers and consumers act as active facilitators and passive followers, respectively. The upper model is based on the maximizing of comprehensive energy application providers' interests, where the price of electricity or heating is determined by the IEO; while the lower model is based on the user selecting energy consumption times based on price to reduce energy consumption costs, with the power consumption plan being given back to the IEO. The schematic diagram of game model is demonstrated in Fig. 2 below.

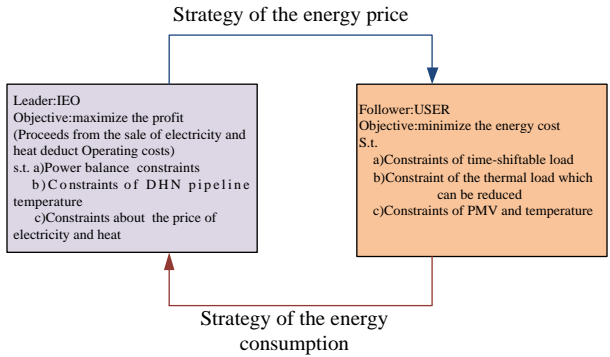

Fig. 2 Schematic diagram of game model

### 3.1 The upper-level model
#### 3.1.1 The objective function of upper-level model

After deducting the total operating cost of the system, the upper objective function is the greatest profit, which may be represented as:

$$\max C = C_{prof} - C_{opf} \quad (15)$$

where $C$ is the profit amount of the integrated energy supplier; $C_{opf}$ is the system operating cost; $C_{prof}$ is the total income for the integrated energy supplier by selling electric energy and heat energy.

**(1) System economic dispatching cost**

The IES's economic dispatch is influenced by the operating costs of the unit, the reserve cost of the units, and the charging and discharging costs of electric storage. The system economic dispatching cost is represented as

$$C_{opf} = \sum_{t=1}^{T}((\sum_{i=1}^{I} F_{CHP}^{t,i} + \sum_{j=1}^{J} F_{CON}^{t,j})\Delta T + \sum_{u=1}^{U} \zeta_u R_{u,t} + \sum_{l=1}^{L} \pi_l (P_{l,cha} + P_{l,dis})) \quad (16)$$

where $T = \{1,2,...,24\}, \forall t \in T$, $T$ is scheduling cycle. $F_{CHP}^{t,i}$ and $F_{CON}^{t,j}$ are the fuel cost of CHP unit $i$ and thermal power unit $j$ at time $t$, $I$ and $J$ are respectively the numbers of CHP and thermal power units, $P_{l,dis}$ and $P_{l,cha}$ are the electric and heat storage device charging and discharging powers, $L$ is the set of electric and heat storage device, $\pi_l$ is the cost factor for charging and discharging, $R_{u,t}$ is the spinning reserve capacity, $\zeta_u$ is the unit and electric storage devices reserve factor, $U$ is the set of CHP units, thermal power units and electric storage devices.

$$F_{CHP}^{t,i} = a_{chp,i}(P_{chp,i,t} + c_{V,i}H_{chp,i,t})^2 + b_{chp,i}(P_{chp,i,t} + c_{V,i}H_{chp,i,t}) + c_{chp,i} \quad (17)$$

$$F_{CON}^{t,j} = a_{con,j}P_{con,j,t}^2 + b_{con,j}P_{con,j,t} + c_{con,j} \quad (18)$$

where $a_{chp,i}$, $b_{chp,i}$ and $c_{chp,i}$ are the fuel cost coefficients of CHP units, and $a_{con,j}$, $b_{con,j}$ and $c_{con,j}$ are the fuel cost coefficients of thermal power units. $P_{chp,i,t}$ and $H_{chp,i,t}$ are respectively electric power and thermal power of CHP unit $i$ at time $t$, $c_{V,i}$ is the thermoelectric ratio of CHP unit $i$, $P_{con,j,t}$ is electric power of the $j$-th thermal power unit at time $t$.

**(2) Profitable amount of integrated energy operators**

The following formula can be used to calculate the profit of IEO:

$$C_{prof} = \kappa_{e,t}P_{L,t} + \kappa_{h,t}(H_{L,res,t} + H_{L,pub,t}) \quad (19)$$

where $C_{prof}$ is the income from selling electricity and heat energy to users, $\kappa_{e,t}$ and $\kappa_{h,t}$ are respectively real-time electricity and heat prices at time $t$. $P_{L,t}$ is the power load at time $t$, $H_{L,res,t}$ and $H_{L,pub,t}$ are heating load of residential building area, public building area at time $t$.

#### 3.1.2 Constraints of upper-level model
**(1) Power balance constraints**

Electricity cannot be significantly stored. It must be created immediately before usage. Each node of the system must continuously maintain an electric energy balance in order to satisfy load demand.

$$\sum_{i=1}^{I} P_{chp,i,t} + \sum_{j=1}^{J} P_{con,j,t} + P_{wind,t} + P_{pv,t} + P_{e,dis,t}\eta_{e,dis} - \quad (20)$$
$$P_{e,cha,t} / \eta_{e,cha} = P_{L,t}$$

The thermal network must also adhere to the following requirements for thermal power balance:

$$\sum_{i=1}^{I} H_{chp,i,t} + P_{eb}\eta_{eb} + P_{h,dis,t}\eta_{h,dis,t} - P_{h,cha,t} / \eta_{h,cha,t} = \quad (21)$$
$$H_{L,res,t} + H_{L,pub,t}$$

$$Q_{load,t} = H_{L,res,t} + H_{L,pub,t} \quad (22)$$

where $P_{wind,t}$ and $P_{pv,t}$ are respectively the wind and photovoltaic power at time $t$, $P_{e,dis,t}$ and $P_{e,cha,t}$ are discharge power and charge power of electric storage devices at time $t$, $\eta_{e,cha}$ and $\eta_{e,dis}$ are the charging and discharging efficiencies of electric storage devices at time $t$. $H_{L,res,t}$ and $H_{L,pub,t}$ are the heating and heat loads in the residential building area and the public building area, respectively.

**(2) Constraint of hot water pipe's temperature**

$$T_i^{\min} <= T_{i,t} <= T_i^{\max} \quad (23)$$

where $T_{i,t}$ is the temperature of the $i$-node of the thermal network at time $t$; and $T_i^{\min}$ and $T_i^{\max}$ are its minimum and maximum values.

**(3) Constraint of the real-time prices**

The pricing range is set as specified in the formula below to keep electricity and heat prices within a reasonable range.

$$\sum_{t=1}^{T} \kappa_{e,t} = \kappa_{e,t,mean}T, \sum_{t=1}^{T} \kappa_{h,t} = \kappa_{h,t,mean}T \quad (24)$$

$$\kappa_{e,\min} \leq \kappa_{e,t} \leq \kappa_{e,\max}, \quad \kappa_{h,\min} \leq \kappa_{h,t} \leq \kappa_{h,\max} \quad (25)$$

**(4) Other general constraints**

The unit output/climbing constraints and energy storage constraints are not described here. More details regarding these operational constraints can be found in previous fundamental works [44-47].

### 3.2 The lower-level model
#### 3.2.1 The objective function of lower-level model

The objective function of the lower layer is to reduce energy costs and the penalty cost of thermal comfort loss.

$$\min F = \sum_{t=1}^{T} [\kappa_{e,t}P_{L,t} + \kappa_{h,t}(H_{L,res,t} + H_{L,pub,t}) + \psi \cdot (H_{cut,r,t}^2 + H_{cut,p,t}^2)]$$

$$= \sum_{t=1}^{T} \kappa_{h,t}(H_{L0,r,t} - H_{cut,r,t} + H_{L0,p,t} - H_{cut,p,t}) + \psi \cdot (H_{cut,r,t}^2 + H_{cut,p,t}^2)]$$
$$(26)$$

where $P_{L0,t}$ is the electric load before cutting at time $t$, $\kappa_{e,t}$ and $\kappa_{h,t}$ are the real time price of electricity and heat energy, respectively, and $\psi$ is the thermal comfort loss coefficient. $H_{cut,r,t}$ and $H_{cut,p,t}$ are thermal comfort losses in residential building areas and public areas, respectively (their values are derived from the amount of thermal load deviation from the most comfortable temperature); and the penalty cost of thermal comfort loss is $\psi \cdot (H_{cut,r,t}^2 + H_{cut,p,t}^2)$.

#### 3.2.2 Constraints of lower-level model
**(1) Constraints of time-shiftable load**

$$\sum_{t=1}^{24} P_t^{L,TS} = \vartheta \sum_{t=1}^{24} P_{L0,t} \;:\; \varepsilon \quad (27)$$

$$P_{t,\min}^{L,TS} \leq P_t^{L,TS} \leq P_{t,\max}^{L,TS} \;:\; \lambda_{1t}, \lambda_{2t} \quad (28)$$

where $P_t^{L,TS}$ is the time-shiftable load at time $t$, $\vartheta$ is the proportion of time-shiftable loads, $\varepsilon$, $\lambda_{1t}$, $\lambda_{2t}$ are the Lagrange multipliers of the equation constraints about the upper and lower limits of the transfer load, respectively.

**(2) Constraints of cuttable thermal loads**

The threshold value that can curtail the heat load is determined by combining the heating indoor temperature limits in residential and public places (such as Sections 1.3.1 and 1.3.2) with Equation (13).

$$0 \leq H_{cut,r,t} \leq H_{L0,r,t} - H_{cut,r,t,\max} : \lambda_{3,t}, \lambda_{4,t} \quad (29)$$
$$0 \leq H_{cut,p,t} \leq H_{L0,p,t} - H_{cut,p,t,\max} : \lambda_{5,t}, \lambda_{6,t} \quad (30)$$

where $H_{cut,r,t}$, $H_{cut,p,t}$, $H_{L0,r,t}$, $H_{L0,p,t}$ are the reduction or initial heating load of residential area and public area at time $t$, respectively. Note that, the initial heating load takes into account the heat load operating under optimal comfortable conditions. $H_{cut,r,t,\max}$, $H_{cut,p,t,\max}$ are the maximum value of the heat load that can be reduced in the two regions at time $t$, respectively. $\lambda_{3,t}$, $\lambda_{4,t}$, $\lambda_{5,t}$, $\lambda_{6,t}$ are the Lagrange multiplier of the upper and lower limits of the thermal load reduction amount.

**(3) Constraints of PMV and temperature**

In order to better accurately represent the range of potential heating loads in residential and public places while taking district heating features into consideration, the heat load index for interior heating is calculated using temperature and PMV values. The heating load of the residential area is set according to the PMV index; while the heating load of the public area is set based on the PMV index during working hours, and it is set according to the indoor temperature not lower than 5 degrees Celsius during non-working hours. The PMV index of the residential area is set according to the

following formula:

$$\begin{cases} |PMV| \leq 0.5 & t \in [7,20] \\ |PMV| \leq 1 & t \in [0,7) \cup (20,24] \end{cases} \quad (31)$$

The PMV index and temperature of the public area are set according to the following formula:

$$\begin{cases} |PMV| \leq 0.5 & t \in [8,22] \\ t_{in} \geq 5℃ & t \in [0,8) \cup (22,24] \end{cases} \quad (32)$$

# 4 Solution of the proposed game-based scheduling model

## 4.1 The framework of the Stackelberg game

The IEO leads the game model, which is followed by the user, resulting in the Stackelberg game model.

$$G = \{N; \phi_{IEO}; \varphi_{USER}; C; F\} \quad (33)$$

where $N$ is participants in the model, $\phi_{IEO}$ and $\varphi_{USER}$ are the strategies adopted by the IEO and user, $C$ and $F$ are objectives of the respective strategies. Among them, the participants include the IEO as the leader of the model and the user as the follower, so it can be expressed as: $N = \{IEO; USER\}$. In this two-layer game model, the operators supply the customers with real-time electricity and heat prices, and its strategy is $\phi_{IEO} = (\kappa_{e,t}, \kappa_{h,t})$. On the user side, the energy consumption plan is changed by the IEO, and take $\varphi_{USER} = (P_t^{L,TS}, H_{cut,r,t}, H_{cut,p,t})$ as its strategy. Leaders and followers develop strategies based on objective functions like equations (15) and (26).

When leaders adopt strategies $x \in \phi_{IEO}$, the user reacts with strategies $y(x) \in \varphi_{USER}$ for them, and leaders respond with new strategies $x(y(x)) \in \phi_{IEO}$ based on this feedback. This cycle is repeated several times until the leader and the user reach the Stackelberg game equilibrium $(x^*, y^*)$, which equals $(\phi_{IEO}^*, \varphi_{USER}^*)$. The following formula is satisfied when the strategy achieves an equilibrium solution:

$$\begin{cases} C(\phi_{IEO}^*, \varphi_{USER}^*) \geq C(\phi_{IEO}, \varphi_{USER}^*) \\ F(\phi_{IEO}^*, \varphi_{USER}^*) \leq F(\phi_{IEO}^*, \varphi_{USER}) \end{cases} \quad (34)$$

In other words, when a solution is found that is agreeable to both the leader and the user, it is the best course of action for both parties, and neither can profit by changing the strategy.

## 4.2 Mathematical analytic formula of the established model

The two-layer optimization problem is broken down in this paper into mathematical programs with equilibrium constraints. The objective function and constraints of the lower model are converted into the upper model's constraints using KKT conditions. Nonlinear constraints are managed using the big-M method, which converts them into a mixed-integer quadratic programming (MIQP) problem. Note that, this work uses probability density functions for modeling the uncertainty of renewables. Due to the difficulty in obtaining precise probability distribution functions of uncertain parameters in real applications, deep generative learning-based renewable scenario generation can be adopted to more realistically model the renewable uncertainties [48].

The description of the Lagrangian function for creating a lower level model is as follows:

$$L(P_t^{L,TS}, H_{cut,r,t}, H_{cut,p,t}, \lambda_{1t}, \lambda_{2t}, \lambda_{3t}, \lambda_{4t}, \lambda_{5t}, \lambda_{6t}) = F +$$
$$\varepsilon(\sum_{t=1}^{24} P_t^{L,TS} - \vartheta \sum_{t=1}^{24} P_{L0,t}) - \lambda_{1t}(P_t^{L,TS} - P_{t,\min}^{L,TS}) -$$
$$\lambda_{2t}(P_{t,\max}^{L,TS} - P_t^{L,TS}) - \lambda_{3t} H_{cut,r,t} -$$
$$\lambda_{4t}(H_{L0,r,t} - H_{cut,r,t,\max} - H_{cut,r,t}) -$$
$$\lambda_{5t} H_{cut,p,t} - \lambda_{6t}(H_{L0,p,t} - H_{cut,p,t,\max} - H_{cut,p,t})$$
$$(35)$$

Taking the partial derivatives of equation (35) respectively, we can get:

$$\frac{\partial L}{\partial P_t^{L,TS}} = \kappa_{e,t} + \varepsilon - \lambda_{1t} + \lambda_{2t} = 0 \quad (36)$$

$$\frac{\partial L}{\partial H_{cut,r,t}} = -\kappa_{h,t} + 2H_{cut,r,t} - \lambda_{3t} + \lambda_{4t} = 0 \quad (37)$$

$$\frac{\partial L}{\partial H_{cut,p,t}} = -\kappa_{h,t} + 2H_{cut,p,t} - \lambda_{5t} + \lambda_{6t} = 0 \quad (38)$$

Therefore, the two-layer optimization problem can be transformed into the following model:

*Max* (15)

*Subject* to (20)-(24), (36)-(38)

$$(P_t^{L,TS} - P_{t,\min}^{L,TS}) \perp \lambda_{1t}, (P_{t,\max}^{L,TS} - P_t^{L,TS}) \perp \lambda_{2t},$$
$$(H_{L0,r,t} - H_{cut,r,t,\max} - H_{cut,r,t}) \perp \lambda_{4t}, \quad (39)$$
$$H_{cut,p,t} \perp \lambda_{5t}, H_{cut,r,t} \perp \lambda_{3t},$$
$$(H_{L0,p,t} - H_{cut,p,t,\max} - H_{cut,p,t}) \perp \lambda_{6t}$$

*where* (31)-(32)

Since the constraints in Eq. (39) are nonlinear, the big-M method is used to linearize them, $(P_{t,\max}^{L,TS} - P_t^{L,TS}) \perp \lambda_{2t}$ can be equivalently transformed into:

$$P_{t,\max}^{L,TS} - P_t^{L,TS} \leq \nu M \quad \lambda_{2t} \leq (1-\nu)M \quad (40)$$

where $M$ is a sufficiently large positive constant, $\nu$ is zero-one variable.

## 4.3 Solution flow

This study addresses the demand response of a

thermal load differential and solves the problem using the master-slave game approach as follows:

Step 1: Establish a probability density function and provide a probabilistic description of the wind and solar outputs.

Step 2: Process the continuous sequence of scenery using the serialization operation.

Step 3: Process the wind and solar output and backup constraints using the opportunity constraint method, and change the uncertainty constraints into deterministic constraints.

Step 4: Establish a power grid flow model, a heat network hydraulic model, and a thermal model by entering pertinent parameters.

Step 5: Determine the upper and lower limits of the residential heating load for each heating period using the PMV index.

Step 6: The heating load threshold value for each time period in the public area is established based on the PMV index during working hours, the temperature during non-working hours meeting the duty temperature.

Step 7: Create a lower layer based on the game model.

Step 8: List the KKT conditions for the lower-level issue.

Step 9: Transform the nonlinear constraints using the big-M method.

Step 10: Create an optimization model for the top-level integrated energy operator.

Step 11: Use YALMIP and the Cplex solver on the MATLAB platform to solve the constructed model.

Step 12: Obtain a two-layer scheduling scheme for optimization, and then output the results.

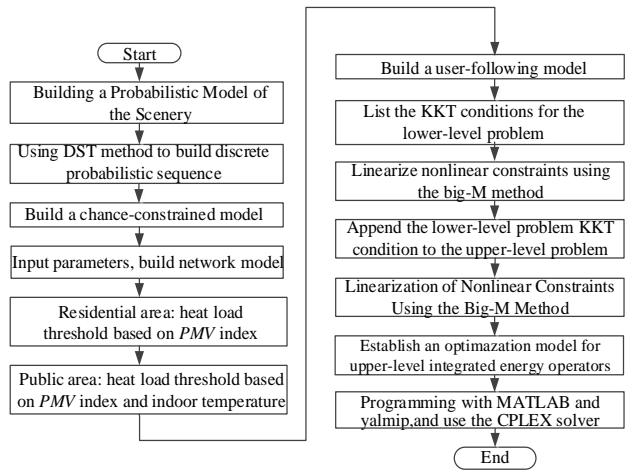

Fig. 3 Solution flow of the proposed method

## 5 Case studies

The test system used in this research combines the IEEE 30-bus power system and two 6-node heat network models [39], as indicated in the Fig. 4. Residential and public areas are heated by the two six-node heat networks, respectively.

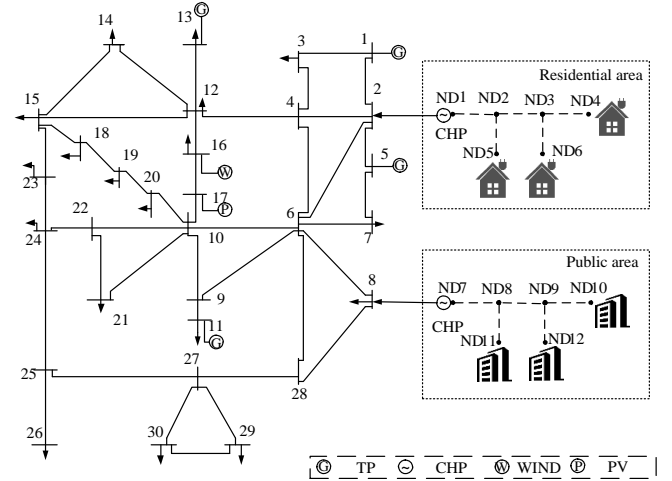

Fig. 4 System structure

In this test system, the IEEE 30-bus power system has 4 thermal power units connected to nodes 1, 5, 11, and 13, two CHP units connected to nodes 2 and 8, and wind and solar power plants installed at nodes 16 and 17. Tables B1 and B2 display the pertinent parameters of thermal power units and CHP units. Fig. 5 displays the ambient temperature, wind speed, solar output, and electric and heat load data prior to the response. Due to without consideration of the secondary heat network, the equivalent heat loads in residential and public areas are respectively aggregate at nodes 4, 5, and 6 and nodes 10, 11, and 12. The time-step size in this study is one hour, and the scheduling cycle is 24 hours.

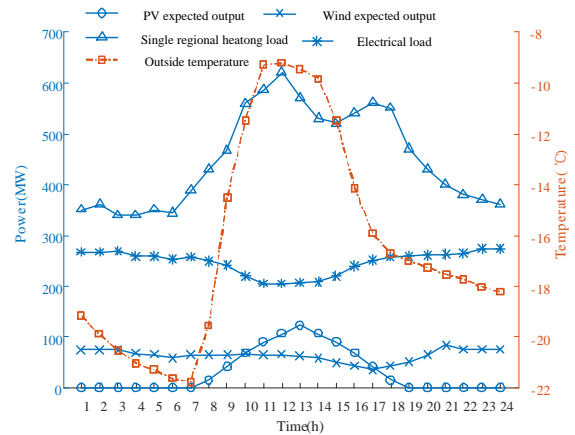

Fig. 5 Data of outdoor temperature, wind and PV expected output and electric, heating load

The parameters of wind, PV, electric storage device and heat storage device are shown in the Table B3. The specific heat capacity of indoor air is $c_{air} = 1.007 kJ \cdot kg^{-1} \cdot {}^\circ C^{-1}$. The density of indoor air is $\rho_{air} = 1.2 kg \cdot m^{-3}$. The total heating volume of the two areas is $6 \times 10^7 m^3$. The proportion of residential area is $K=0.5$. The comprehensive heat transfer coefficient of the building is $K_{htc} = 0.5 W \cdot m^{-2} \cdot {}^\circ C^{-1}$.

The heating volume of the two areas can be calculated from the total heating volume and the proportion of residential area. Then, two areas' heating acreage is calculated using the building shape factor. The maximum and minimum water supply temperatures for the heating network are 100°C and 90°C, respectively. The maximum and minimum return water temperature is 60°C and 35°C. Set the appropriate real-time price parameters, such as 90/40 $/MWh for the maximum/minimum electricity price, 65$/MWh for the average, and 70/30 $/MWh for the maximum/minimum hot price, 50$/MWh for the average.

The energy consumption time and building characteristics are directly connected to the heat load requirement required by various functional regions. In order to meet users' needs for thermal comfort, a system is also offered that takes time-sharing and partitioning into consideration when calculating heat load requirements. Other modes are configured to contrast it at the same time. The following modes are designed in this study based on the flexibility and effect of the obtained strategy to reduce heat load:

Mode 1: The indoor temperature in both the residential and public areas maintains within the required thermal comfort range at all times.

Mode 2: The temperature in both residential and public places is set to the ideal level.

Mode 3: The temperature in the residential area is kept at the required thermal comfort range, while the public area remains at an ideal temperature.

Mode 4: The temperature in the residential area is kept at an ideal temperature, while the public area remains at the required thermal comfort range

Mode 5: The temperature in the residential area is kept at the required thermal comfort range, while the public area adopts different actions during working and non-working hours.

The MATLAB software is used in this simulation, along with the YALMIP plug-in to invoke the CPLEX solver to solve the problem.

## 5.1 Economic analysis

To analyze the economy of the proposed method, the IEO benefits and user costs are studied under the five modes. Table 1 lists the three indicators of operation cost, income and net income of IEO, and three indicators of user-side energy consumption cost and comfort loss, and total cost. The comparative analysis of Modes 1-5 shows that: in Mode 1, the user-side energy cost is the largest, and the IEO benefits are the largest. Mode 2 operates at the optimum temperature with no loss of thermal comfort. The net income of the IEO of Mode 5 is at the middle level, the decrease in net income accounts for 3.5% of the net income of Mode 1, the increase accounts for 2.1% of Mode 3, and the operating cost is at a lower level. Mode 3 is small, the energy cost is the minimum value in the proposed strategy, the comfort loss is the largest, and the total user cost is at a low level, which is 36,617$ less than Mode 1. The above analysis demonstrates that due to the different heat load requirements of different functional areas, it is advantageous to set the range of heat load response based on the energy consumption characteristics of the area. It also can be found that the resulting scheduling strategy can significantly improve IEO's revenue and reduce user costs at the expense of user thermal comfort to a certain extent. Note that, the loss of thermal comfort is still tolerable.

Table. 1 Revenue and user cost analysis of integrated energy applications

| Mode | IEO | | | User | | |
|---|---|---|---|---|---|---|
| | Net revenue/$ | Earnings/$ | Operating costs/$ | Energy costs/$ | The cost of comfort loss /$ | Overall costs/$ |
| 1 | 957709.2 | 1123614 | 156230.8 | 1123614 | 10196.8 | 1111134 |
| 2 | 935227.2 | 1099309 | 154635.3 | 1099309 | / | 1099309 |
| 3 | 904993.1 | 1065153 | 151018.7 | 1065153 | 9383.6 | 1053046 |
| 4 | 926553.9 | 1075803 | 149249.3 | 1075803 | 6271.0 | 1082074 |
| 5 | **924042.6** | 1054377 | 150474.1 | 1054377 | 20139.7 | **1074517** |

## 5.2 Analysis of unit outputs

Fig. 6 depicts the electric and thermal output of CHP units in five different modes. According to Fig. 6, there are significant disparities in the heat output of CHP units, with CHP units in mode 5 having the lowest heat output. The electric outputs of CHP units in five modes is relatively flat. This result demonstrates that the proposed technique has considerable advantages in terms of reducing CHP unit heat production and fossil energy use.

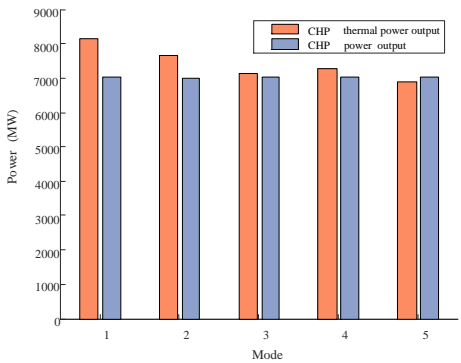

Fig. 6 Electric and thermal outputs of CHP units of different modes

## 5.3 Analysis of integrated demand response
### 5.3.1 Time-shiftable load

As a controllable load, the time-shiftable load will shift the energy consumption period under the incentive of real-time electricity price and under the premise that the total electricity consumption remains unchanged. According to the price of electricity, users can find the most economical time to use electricity. Peak and valley filling can reduce load pressure during peak load hours, move energy load to load valleys. The schematic diagram of the change of electrical load before and after demand response is shown in Fig. 7. The peak load was lower between the hours of 9:00 and 19:00. When the electrical load rises, which also happens to be a period when a lot of wind energy is produced, wind power curtailment is somewhat less.

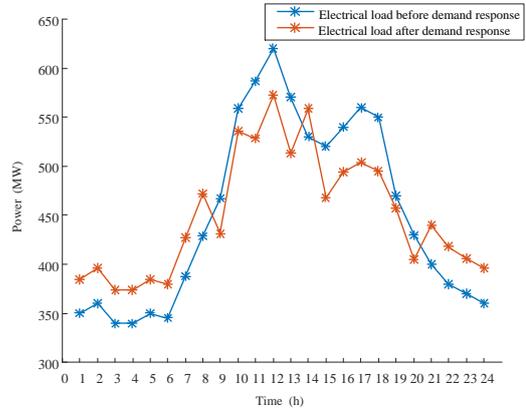

Fig. 7 Change of electrical load before and after demand response

### 5.3.2 Cuttable heat load

As shown in Fig. 8, the thermal load of public and residential places is determined by human comfort in mode 1. It can be observed that while the heating price is low, the heat load can be lowered to a little extent, and when the heating price is high, the heat load may be reduced to a big extent. In Mode 5, the cost of heating is inversely proportional to the amount of heat load that can be lowered. Compared with Mode 1, the technique described in this work (Mode 5) can cut more heat load, and help the system spend less energy. In Fig. 9, Mode 5 outperforms Modes 1, 3 and 4 with the maximum heat load reduction. Modes 3 and 4 maintain optimal temperatures in residential and public areas, respectively; while Modes 1 and 5 have a relatively large temperature margin for adjustment during 00:00-7:00 and 22:00-24:00.

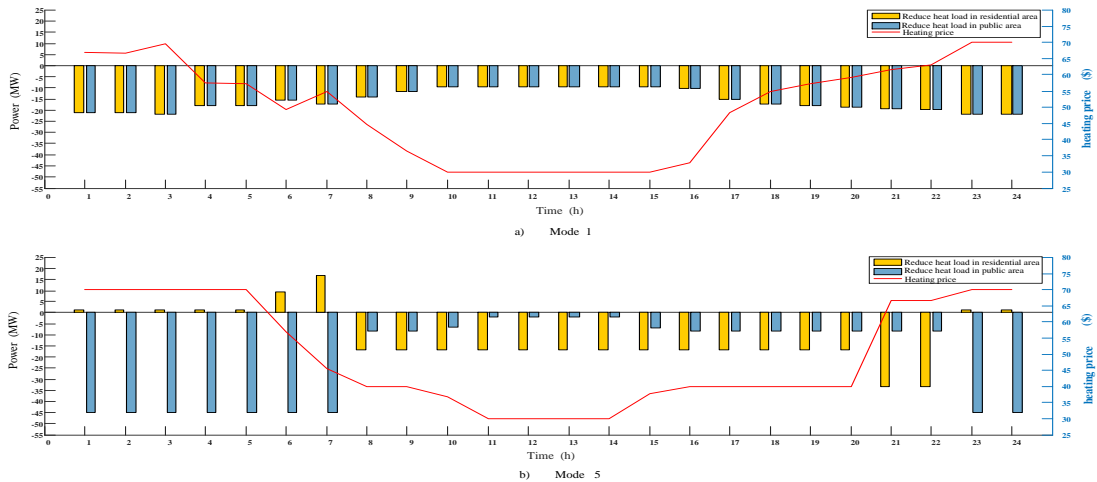

Fig. 8 Comparison and real-time heat price of reducible heat load between Mode 1 and Mode 5

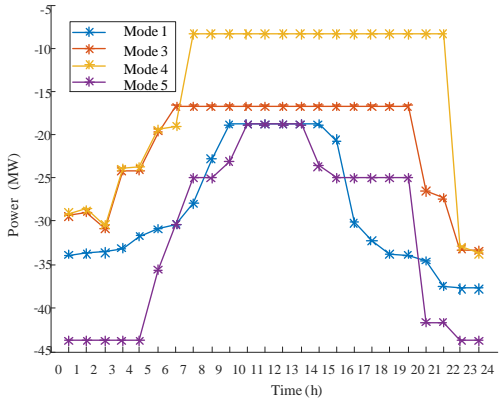

Fig. 9 Thermal load reduction of modes 1, 3, 4 and 5

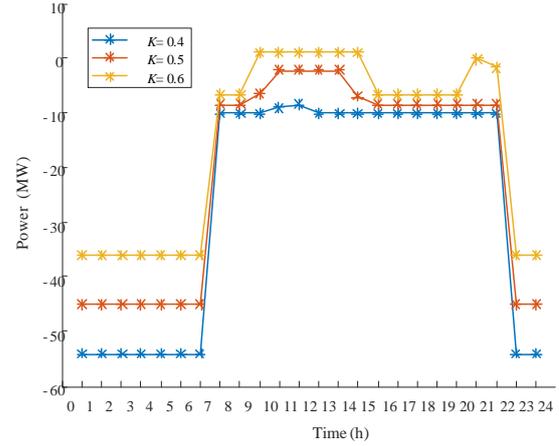

Fig. 10 Analysis of total heat reduction versus residential areas

## 5.4 Sensitivity analysis

As shown in Fig. 10, $K$ is the proportion of residential heating volume to total heating volume, and $K=0.4$, $K=0.5$, and $K=0.6$ are chosen for analysis. Among them, when $K=0.5$, the heating volume of the residential area is equal to the public area; when $K=0.4$, the proportion of the residential area is lower than that of the public area; when $K=0.6$, the opposite is true. Fig. 10 shows the sensitivity curve that can reduce the total heat load and the proportion of residential area. The three curves in the figure all have large reductions in the periods of 0:00-7:00 and 23:00-24:00. However, it can be seen that when the ratio is 0.4, the heat load can be reduced more. When planning new areas, public areas can be increased within a certain margin zone ratio for more heat load flexibility.

Table 2 shows the sensitivity analysis results of changing the proportion of residential areas. From Table 2, the following results can be obtained:

(1) With the increase of the proportion of the residential area, the energy consumption cost of the user is increasing, and the comfort loss is decreasing. Since cuttable heat loads increase during non-working hours, the electrical output of the thermoelectric coupling is reduced. As a result, the cost of generating energy and operating costs of the unit are relatively low.

(2) The income of the IEO increases with the proportion due to the increase of the energy consumption cost on the user side. The net income of the IEO is the largest when the proportion is the largest. The total cost of the user is smaller when the proportion is smaller, and the most beneficial solution for the user is obtained. It can be concluded that when the proportion is close to the median value, the interests of IEO and users can be balanced.

Table. 2 Sensitivity analysis of residential area proportions

| | IEO | | | User | | |
|---|---|---|---|---|---|---|
| $K$ | Net revenue /$ | Earnings /$ | Operating costs /$ | Energy costs /$ | The cost of comfort loss /$ | Overall costs/$ |
| 0.1 | 854678.1 | 947419.4 | 143194 | 947419.4 | 50452.68 | 997872.1 |
| 0.2 | 868351.2 | 978642.1 | 150078.1 | 978642.1 | 39787.17 | 1018429 |
| 0.3 | 899139.2 | 1011068 | 146347.5 | 1011068 | 34418.82 | 1045487 |
| 0.4 | 916758.9 | 1042311 | 152975.6 | 1042311 | 27423.95 | 1069734 |
| 0.5 | 924042.6 | 1054377 | 150474.1 | 1054377 | 20139.69 | 1074517 |
| 0.6 | 948115.8 | 1087124 | 155925.5 | 1087124 | 16917.75 | 1104041 |
| 0.7 | 995421.6 | 1127939 | 156325.7 | 1127939 | 23808.32 | 995421.6 |
| 0.8 | 1043279 | 1170367 | 155358.2 | 1170367 | 28269.84 | 1198637 |
| 0.9 | 1116348 | 1244473 | 163343.8 | 1244473 | 35218.55 | 1279691 |

## 5.5 Comparative analysis of master-slave game model and single-layer model

To examine the performance of the presented master-slave game model, a comparative analysis of the proposed model and single-layer model has been carried out. As for the master-slave game model, the results of Mode 5 are used for subsequent analysis. Regarding the single-layer model, the IEO only seeks to the minimum energy costs without consideration of user costs; an electric or heating pricing strategy based on time-of-use prices is adopted to guide users' behavior, and the price is allocated according to the percentage of the load in the overall amount; the user comfort cost is set as the penalty cost of thermal comfort loss in Eq. (26). Expect the price constraints, the both models (master-slave game model and single-layer model) have the same constraints. The PMV index is used to estimate the interior heat load of users in residential areas, whereas the PMV index and the interior temperature are used to determine the interior heat load of users in public places, according to Mode 6. The objective function in an independent operation mode is as follows:

$$C = \sum_{t=1}^{T}((\sum_{n=1}^{N}F_{CHP}^{t,n} + \sum_{m=1}^{M}F_{CON}^{t,m})\Delta T + \sum_{k=1}^{K}\zeta_k R_{k,t} + \sum_{l=1}^{L}\pi_l(P_{l,cha} + P_{l,dis})) - \kappa_{e,t}P_{L,t} - \kappa_{h,t}(H_{L,res,t} + H_{L,pub,t})$$

(41)

Tables 3 and 4 show the cost comparison, economic comparison and heat comparison analysis between the two-layer model and the single-layer model considering the master-slave game. It can be seen from the above table that:

(1) Compared with the single-layer model, the master-slave game model has the higher thermal comfort penalty cost and the lower energy cost and operating cost. The difference in the thermal comfort penalty cost between the two models is not reflected in the heat reduction, and master-slave game model has a greater heat reduction than single-layer model. The decrease in energy costs reflects the fact that the lower heat load contributes to a decrease in energy consumption. At the same time, along with the decline in energy costs, operating costs have been reduced. The results show that the two-layer model of the master-slave game has a better guiding effect on reducing the heat load than the single-layer model. The two-tier model also plays a role in reducing energy costs and operating costs.

(2) The users' cost and the benefits of IEO in single-layer model are higher those in master-slave game model. In comparison to master-slave game model, single-layer model has higher user costs and supplier benefits. This is because the Stackelberg game strategy can balance the interests of suppliers and users, but the single-layer model only seeks to minimize the supplier's cost or to maximize the supplier's profit, and pays less attention to the user's side interests. The advantages of the Stackelberg game are reflected, and it is concluded that the two-layer model is more likely to obtain a satisfactory equilibrium solution than the single-layer model.

Table. 3 Comparison of mode 5 and mode 6 related indicators

| Master-slave game two-layer model (Mode 5) | | | Single Layer Model (Mode 6) | | |
| --- | --- | --- | --- | --- | --- |
| Operating costs/$ | Energy cost/$ | Thermal Comfort Penalty Cost/$ | Operating costs/$ | Energy cost/$ | Thermal Comfort Penalty Cost/$ |
| 150474.1 | 1054377 | 20139.7 | 178063.8 | 1154759 | 12711.4 |

Table. 4 Comparison of mode 5 and mode 6 economy and thermal output

| Master-slave game two-layer model (Mode 5) | | | | Single Layer Model (Mode 6) | | | |
| --- | --- | --- | --- | --- | --- | --- | --- |
| Supplier revenue/$ | user cost/$ | Total reduced heat load/MW | CHP unit heat output/MW | Supplier revenue/$ | user cost/$ | Total reduced heat load/MW | CHP unit heat output/MW |
| 924042.6 | 1074517 | 750 | 6899.1 | 976695.2 | 1167470 | 674.7 | 8323.8 |

(3) To protect the interests of suppliers, the total heat load reduction of single-layer model is less than that of master-slave game model. Only more sales of thermal energy can be more profitable for IEO. Taking into account the differences in heat demand across multiple regions, the strategy suggested in this paper is able to significantly lessen the impact of district heating over time. Compared with Mode 6, the heat produced by CHP units in Mode 5 is less, which increases the amount of wind power that can be transferred to the main grid in Mode 5. It can be safely drawn that the proposed method based on the Stackelberg master-slave game is better than the single-layer model in this paper.

# 6 Conclusion

In this paper, various demand-side management strategies for heat load are implemented in accordance with the building shape characteristics, user energy consumption characteristics of public and residential areas, and demand differences of heat load. An IDR model for the Stackelberg game is developed, and it is driven by current electricity and heat prices. A network model based on IEEE30 nodes and two 6-node thermal networks is used to validate the conclusion of this paper. The following conclusions are drawn from the example analysis:

(1) The interests between the IEO and the user reach a Nash equilibrium due to the use of Stackelberg game in the proposed approach, which also lowers the user's energy consumption costs. In comparison to the single-layer model, the benefits of the suggested model in balancing the interests of IEO and users are fully reflected, and the user side's enthusiasm is fully mobilized.

(2) According to the building characteristics and energy consumption characteristics of different areas, the presented approach is advantageous to carry out heat demand response strategies for different areas. It provides margin for thermal load reduction, thereby increasing thermal load flexibility.

(3) Through the sensitivity analysis of the proportion of residential areas, it can be seen that with the increase of the proportion of public areas, the heat that can be reduced by the system will increase, and the efficiency of system energy utilization will be further improved. At the same time, it can be more energy-efficient to match the appropriate proportion of the residential area in the area.

## Appendix A:

### (1) Proof of existence

Theorem: Only when the model satisfies the following three conditions at the same time, the Stackelberg game problem can obtain an equilibrium solution.

(a) The IEO's objective function is the non-empty continuous function of $\phi_{IEO} = (\kappa_{e,t}, \kappa_{h,t})$;

(b) The objective function of USER is a non-empty continuous function of $\varphi_{USER} = (P_t^{L,TS}, H_{cut,r,t}, H_{cut,p,t})$;

(c) The objective function of USER is the quasi-convex function of $\varphi_{USER} = (P_t^{L,TS}, H_{cut,r,t}, H_{cut,p,t})$.

Proof. $C$ and $F$ are nonempty and continuous functions of $(\kappa_{e,t}, \kappa_{h,t})$ and $(P_t^{L,TS}, H_{cut,r,t}, H_{cut,p,t})$, respectively. Thus, the proof for condition (c) is as follows:

$$\frac{\partial^2 F}{\partial (P_t^{L,TS})^2} = 0 \quad (1)$$

$$\frac{\partial^2 F}{\partial (H_{cut,r,t})^2} = 2\psi > 0 \quad (2)$$

$$\frac{\partial^2 F}{\partial (H_{cut,p,t})^2} = 2\psi > 0 \quad (3)$$

From equations (1)-(3), it can be concluded that $F$ is a convex function about $(P_t^{L,TS}, H_{cut,r,t}, H_{cut,p,t})$. Therefore, the existence of the equilibrium solution of the established game model is proved.

### (2) Proof of uniqueness

Theorem: When the Stackelberg game satisfies the following conditions, a unique equilibrium solution exists.

(a) When the IEO's strategy is defined as $\phi_{IEO} = (\kappa_{e,t}, \kappa_{h,t})$, the USER take $\varphi_{USER} = (P_t^{L,TS}, H_{cut,r,t}, H_{cut,p,t})$ as the unique strategy;

(b) When the USER's strategy is defined as $\varphi_{USER} = (P_t^{L,TS}, H_{cut,r,t}, H_{cut,p,t})$, the IEO has a unique optimal solution.

Proof: (a) Compute the first-order partial derivative of $F$ with respect to $P_t^{L,TS}, H_{cut,r,t}, H_{cut,p,t}$, respectively.

$$\frac{\partial F}{\partial (P_t^{L,TS})} = \kappa_{e,t} > 0 \quad (4)$$

$$\frac{\partial F}{\partial (H_{cut,r,t})} = -\kappa_{h,t} + 2\psi H_{cut,r,t} \quad (5)$$

$$\frac{\partial F}{\partial (H_{cut,p,t})} = -\kappa_{h,t} + 2\psi H_{cut,p,t} \quad (6)$$

According to formula (4), it can be known that the function of $F$ is an increasing trend according to the variable $P_t^{L,TS}$. Let the first-order partial derivatives in equations (5) and (6) be equal to zero, respectively. Simplified to: $H_{cut,r,t} = \kappa_{h,t}/2\psi$ and $H_{cut,p,t} = \kappa_{h,t}/2\psi$. When $\kappa_{h,t}/2\psi < 0$, the optimal strategy of users is $(P_t^{L,TS}, 0, 0)$. When $0 \leq \kappa_{h,t}/2\psi \leq H_{L0,r,t} - H_{cut,t,\max}$, the optimal strategy of users is $(P_t^{L,TS}, \kappa_{h,t}/2\psi, \kappa_{h,t}/2\psi)$. When $\kappa_{h,t}/2\psi \geq H_{L0,r,t} - H_{cut,t,\max}$, the optimal strategy of users is $(P_t^{L,TS}, H_{L0,r,t} - H_{cut,r,t,\max}, H_{L0,r,t} - H_{cut,p,t,\max})$.

(b) The derivatives of $C$ with respect to $\kappa_{e,t}$ and $\kappa_{h,t}$

respectively are:

$$\frac{\partial C}{\partial (\kappa_{e,t})} = P_{L,t} > 0 \quad (7)$$

$$\frac{\partial C}{\partial (\kappa_{h,t})} = H_{L,res,t} + H_{L,pub,t} > 0 \quad (8)$$

It can be known from equations (7)-(8) that $C$ is a monotonically increasing function of $\kappa_{e,t}$ and $\kappa_{h,t}$. Therefore, the revenue model of IEO can obtain a unique solution about $\kappa_{e,t}$ and $\kappa_{h,t}$.

**Appendix B:**

Table. B1 Main parameters of thermal power units

| $P_{con,i,max}$ (MW) | $P_{con,i,min}$ (MW) | $r_{u,j}^{con}$ (MW/h) | $r_{d,j}^{con}$ (MW/h) | $a_{con,j}$ ($/MW²) | $b_{con,j}$ ($/MW) | $c_{con,j}$ ($) | $\zeta_u$ ($/MW) |
|---|---|---|---|---|---|---|---|
| 50 | 25 | 25 | 25 | 0.012 | 17.82 | 10.150 | 13.7 |
| 35 | 10 | 18 | 18 | 0.069 | 26.24 | 31.670 | 13.2 |
| 30 | 10 | 15 | 15 | 0.028 | 37.69 | 17.940 | 13.2 |
| 40 | 12 | 20 | 20 | 0.010 | 12.88 | 6.778 | 14.2 |

Table. B2 Main parameters of CHP units

| $P_{chp,i,max}$ (MW) | $P_{chp,i,min}$ (MW) | $H_{chp,i,max}$ (MW) | $r_{u,i}^{chp}$ (MW/h) | $r_{d,i}^{chp}$ (MW/h) | $a_{chp,i}$ ($/MWh) | $b_{chp,i}$ ($/MWh) | $c_{chp,i}$ ($/h) | $C_V$ | $\zeta_u$ ($/MW) |
|---|---|---|---|---|---|---|---|---|---|
| 200 | 100 | 250 | 50 | 50 | 0.0044 | 13.29 | 39 | 0.15 | 16.2 |
| 200 | 100 | 250 | 50 | 50 | 0.0044 | 13.29 | 39 | 0.15 | 16.2 |

Table. B3 Parameters of wind turbines and PV

| parameter | data |
|---|---|
| Wind turbines parameters | $v_{in}$ =3m/s |
| | $v_r$ =15m/s |
| | $v_{out}$ =25m/s |
| | $P_r$ =60MW |
| Photovoltaic parameters | $A_{pv}$ =1300m² |
| | $P_{max}^{PV}$ =120MW |

Table. B4 Parameters of energy storage units

| parameter | data |
|---|---|
| Electric storage device parameters | $P_{max}^{DC} = P_{max}^{CH}$ =40MW |
| | $S_{oc,min}$ =32MWh |
| | $S_{oc,max}$ =160MWh |
| | $\eta_{dc} = \eta_{ch}$ =0.9 |
| Heat storage device parameters | $P_{max}^{HC} = P_{max}^{HD}$ =50MW |
| | $C_{oc,min}$ =40MWh |
| | $C_{oc,max}$ =240MWh |